\documentclass[twocolumn,prd,floatfix,preprintnumbers,nofootinbib,superscriptaddress]{revtex4-2}

\usepackage{bm}
\usepackage{times}
\usepackage{amssymb,amsbsy,amsmath,amsfonts}
\usepackage{graphicx}
\usepackage{float}
\usepackage{color}
\usepackage{morefloats}
\usepackage{rotating}
\usepackage{srcltx}
\usepackage{slashed}
\usepackage{subfigure}
\usepackage{multirow,diagbox}
\usepackage{verbatim}
\usepackage{tabularx}
\usepackage{tabularray}
\usepackage[outdir=./]{epstopdf}
\usepackage[normalem]{ulem}
\usepackage{hyperref}
\usepackage{tikz}
\hypersetup{colorlinks=true,linkcolor=blue,citecolor=blue,menucolor=black,urlcolor=black,filecolor=blue}

\begin{document}

\title{Shedding light on the nature of $\phi(2170)$ state in the $e^+e^- \to \phi \pi^+\pi^-$ reaction}

\date{\today}

\author{Xiang Wei}
\affiliation{State Key Laboratory of Heavy Ion Science and Technology, Institute of Modern Physics, Chinese Academy of Sciences, Lanzhou 730000, China} \affiliation{School of Nuclear Sciences and Technology, University of Chinese Academy of Sciences, Beijing 101408, China}

\author{Qing-Hua Shen}
\affiliation{State Key Laboratory of Heavy Ion Science and Technology, Institute of Modern Physics, Chinese Academy of Sciences, Lanzhou 730000, China}  \affiliation{School of Physical Science and Technology, Lanzhou University, Lanzhou 730000, China} \affiliation{School of Nuclear Sciences and Technology, University of Chinese Academy of Sciences, Beijing 101408, China}

\author{Xiao-Hai Liu}~\email{xiaohai.liu@tju.edu.cn}
\affiliation{Center for Joint Quantum Studies and Department of Physics, School of Science, Tianjin University, Tianjin 300350, China}

\author{Ju-Jun Xie}~\email{xiejujun@impcas.ac.cn}
\affiliation{State Key Laboratory of Heavy Ion Science and Technology, Institute of Modern Physics, Chinese Academy of Sciences, Lanzhou 730000, China} \affiliation{School of Nuclear Sciences and Technology, University of Chinese Academy of Sciences, Beijing 101408, China} 
\affiliation{Southern Center for Nuclear-Science Theory (SCNT), Institute of Modern Physics, Chinese Academy of Sciences, Huizhou 516000, China}

\begin{abstract}

We investigate the production of $\phi(2170)$ state in the $e^+e^- \to \phi \pi^+\pi^-$ reaction with the effective Lagrangian approach. In addition to the tree level contributions from the $\phi(1680)$ meson and a possible X(1750) state, we consider also the $K_1$-$K$-$\bar{K}$ intermediate state process from the perspective of triangular singularity. Based on the one-photon exchange approximation, a pair of $K_1 \bar{K}$ mesons was firstly produced, and then the $K_1$ meson subsequently decays into $\phi$ and $K$, and the $K\bar{K}$ pair produce the $\pi^+ \pi^-$ through the final state interactions, in which the scalar meson $f_0(980)$ is dynamically generated. We show that the inclusion of the triangle loop diagrams leads to a good description of the new BESIII measurements, especially for the structure of $\phi(2170)$. This provides a novel interpretation of the $\phi(2170)$ state, offering new insights into its fundamental nature which is still unclear. Furthermore, it is found that these measurements on the $e^+e^- \to \phi \pi^+\pi^-$ reaction can be used to determine some of the properties of two $K_1$ mesons with masses around 1610 MeV and 1895 MeV, which are crucial to reproduce the experimental data.

\end{abstract}

\maketitle

\section{Introduction} \label{sec:introduction}

The $\phi(2170)$ state [previously refereed to as the $Y(2175)$] with quantum numbers $I^G(J^{PC}) = 0^-(1^{--})$ was firstly observed in the $\phi f_0(980)$ channel through the initial state radiation (ISR) $e^+e^- \to \gamma_{ISR} \phi f_0(980)$ reaction by BaBar Collaboration in 2006~\cite{BaBar:2006gsq}. This observation was subsequently confirmed by other experimental groups, including the Belle Collaboration~\cite{Belle:2008kuo}, BES and BESIII Collaborations~\cite{BES:2007sqy,BESIII:2014ybv,BESIII:2017qkh,BESIII:2020gnc,BESIII:2021bjn,BESIII:2021aet}. Additionally, further confirmation was provided by the BaBar Collaboration in their subsequent analyses~\cite{BaBar:2007ptr,BaBar:2011btv}. Although the existence of the $\phi(2170)$ state has been firmly established through these above experimental investigations, significant discrepancies persist in the determination of its Breit-Wigner parameters~\cite{Shen:2009mr}, especially concerning the total decay width, as reported in the latest review by the Particle Data Group (PDG)~\cite{ParticleDataGroup:2024cfk}. The internal constituents of the $\phi(2170)$ state are still unknown, which has stimulated extensive theoretical discussions. It is currently one of the most interesting particles in light hadron spectroscopy.

\begin{figure}[htbp]
\centering 
\includegraphics[scale=0.35]{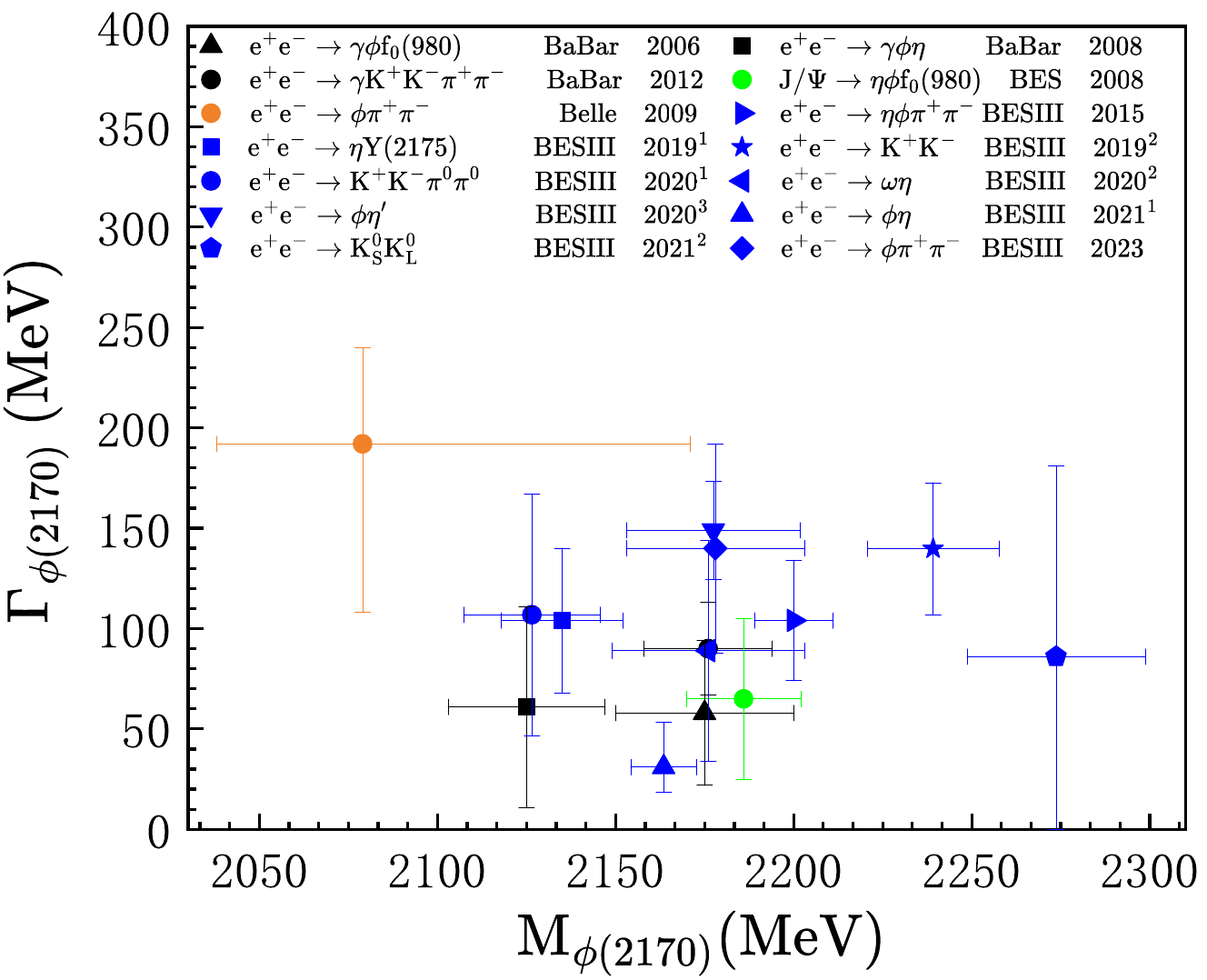}
\caption{A comprehensive compilation of the experimentally determined Breit-Wigner mass and width parameters for the $\phi(2170)$ resonance, where the experimental data are taken from: BaBar 2006~\cite{BaBar:2006gsq}, BaBar 2008~\cite{BaBar:2007ceh}, BES 2008~\cite{BES:2007sqy}, Belle 2009~\cite{Belle:2008kuo}, BaBar 2012~\cite{BaBar:2011btv}, BESIII 2015~\cite{BESIII:2014ybv}, BESIII 2019$^1$~\cite{BESIII:2017qkh}, BESIII 2019$^2$~\cite{BESIII:2018ldc}, BESIII 2020$^1$~\cite{BESIII:2020vtu}, BESIII 2020$^2$~\cite{BESIII:2020xmw}, BESIII 2020$^3$~\cite{BESIII:2020gnc}, BESIII 2021$^1$~\cite{BESIII:2021bjn},BESIII 2021$^2$~\cite{BESIII:2021yam}, and BESIII 2023~\cite{BESIII:2021aet}.} \label{phi2170MW} 
\end{figure}

In Fig.~\ref{phi2170MW}, we collect the experimental results of the resonance parameters, Breit-Wigner mass $M_{\phi(2170)}$ and width $\Gamma_{\phi(2170)}$, of $\phi(2170)$ state obtained from different experimental processes, where one can find that the measured mass and width are different~\cite{Huang:2020ocn}. Especially, the values extracted from the $e^+e^-\rightarrow\phi\pi^+\pi^-$ process by the Belle Collaboration~\cite{Belle:2008kuo} and from the $e^+ e^- \to K^+ K^-$~\cite{BESIII:2018ldc} and $e^+ e^- \to K^0_S K^0_L$~\cite{BESIII:2021yam} reactions by the BESIII Collaboration are much different with others. These observed structures may potentially represent distinct hadronic states, rather than the $\phi(2170)$ resonance. The situation cannot be clarified without further experimental measurements. And the latest average values of the resonance parameters of the $\phi(2170)$ resonance as quoted in PDG~\cite{ParticleDataGroup:2024cfk} are:
\begin{eqnarray}
M_{\phi(2170)} &=& 2164 \pm 6 ~{\rm MeV}, \\
\Gamma_{\phi(2170)} &=& 106^{+24}_{-18} ~ {\rm MeV}.
\end{eqnarray}

On the theoretical side, some explanations for the $\phi(2170)$ sate were proposed, including traditional excited $s\bar{s}$ state~\cite{Ding:2007pc,Wang:2012wa,Afonin:2014nya,Pang:2019ttv,Zhao:2019syt,Li:2020xzs}, $s\bar{s}g$ hybrid state~\cite{Ding:2006ya,Ho:2019org}, $ss\bar{s}\bar{s}$ tetra-quark state~\cite{Wang:2006ri,Chen:2008ej,Drenska:2008gr,Deng:2010zzd,Ke:2018evd,Agaev:2019coa,Liu:2020lpw}, and $\Lambda\bar{\Lambda}$ bound state~\cite{Zhao:2013ffn,Deng:2013aca,Dong:2017rmg}. Moreover, the $\phi(2170)$ state has been extensively studied within the context of the $\phi K\bar{K}$ system with subsystem $K\bar{K}$ as the $f_0(980)$~\cite{MartinezTorres:2008gy,PhysRevD.79.034018}. In this framework, the $f_0(980)$ meson combines with the $\phi$ meson to form a resonance with a mass around 2150 MeV. This dynamically generated state in the $\phi K \bar{K}$ system with quantum numbers $J^{PC} = 1^{--}$, predominantly decays into $\phi f_0(980)$. This decay pattern strongly supports its identification with the $\phi(2170)$ state~\cite{MartinezTorres:2008gy}. Within this picture, the decay modes of $\phi(2170)$ to $\phi \eta$ and $\phi \eta'$ were investigated in Ref.~\cite{Malabarba:2023zez}. Additionally, the strong decays of $\phi(2170)$ to final states involving various kaonic resonances have been conducted in Ref.~\cite{Malabarba:2020grf}, providing valuable insights into the decay dynamics and properties of the $\phi(2170)$ state.

The BESIII results~\cite{BESIII:2020gnc,BESIII:2021bjn,BESIII:2021aet} suggest a nontrivial internal structure for the $\phi(2170)$ state, making it a strong candidate for exotic interpretations, rather than as a simple radial excitation of the vector meson $\phi(1020)$~\cite{Jafarzade:2025qvx}. Nevertheless, it should be noted that the predicted width of the $\phi(2170)$ state from various theoretical models exhibits significant discrepancies when compared with experimental measurements. Therefore, from both experimental and theoretical perspectives, a more precise determination of the $\phi(2170)$ resonance parameters is necessary for a better understanding of its nature.

In this work, based on the recent measurements of the BESIII Collaboration~\cite{BESIII:2021aet}, we advocate a very different picture for the production of the $\phi(2170)$ state. The picture is that the $\phi(2170)$ state is produced in the $e^+ e^- \to \phi \pi^+ \pi^-$ process~\cite{BESIII:2021aet} by the triangular singularity~\cite{Liu:2015taa,Bayar:2016ftu,Guo:2019twa} of the $K_1$-$K$-$\bar{K}$ triangle diagrams, where $K_1$ decays into $\phi$ and $K$ mesons, and then $K$ and $\bar{K}$ can fuse to produce the $\pi^+\pi^-$ through $S$-wave final state interactions, where the scalar meson $f_0(980)$ is dynamically generated~\cite{Oller:1997ti,Oller:1998zr,Gamermann:2006nm}. We will show that the diagram develops a triangle singularity for a $e^+ e^-$ center-of-mass energy of about 2170 MeV when the mass of $K_1$ is about 1610 MeV, and the effects of the triangle singularity shown up clearly in the total cross section around 2170 MeV. Within this picture, the $\phi(2170)$ state was not shown in the $e^+ e^- \to \phi K^+ K^-$ process~~\cite{BESIII:2019ebn} can be also easily understood, because a triangle singularity is strictly constrained by specific energy-dependent conditions, which must be precisely satisfied for its occurrence. Meanwhile, a resonance-like structure, $R(2400)$ might exist around 2.4 GeV in the $e^+ e^- \to \phi \pi^+ \pi^-$ process~\cite{BESIII:2021aet} and it has been also studied by Belle Collaboration~\cite{Belle:2008kuo}, now can be also easily reproduced with a different mass of $K_1$ meson. Thus the $R(2400)$ state can be considered as a partner state of the $\phi(2170)$ state~\cite{Chen:2018kuu}.

In addition, the $\phi(1680)$ resonance also has significant contributions to the $e^+ e^- \to \phi \pi^+\pi^-$ reaction~\cite{BESIII:2021aet}. Thus, its contribution is also taken into account in the tree level. Besides, the BESIII Collaboration~\cite{BESIII:2019dme} has made an important observation that the $X(1750)$ 
resonance~\cite{OmegaPhoton:1984eqn,Busenitz:1989gq,FOCUS:2002ziy,Lichard:2025hxm} appears together with $\phi(1680)$ in decays of $\psi(3686) \to K^+ K^- \eta$. It is shown in the invariant $K^+K^-$ mass spectrum. The $X(1750)$ state is determined to be a $1^{--}$ resonance~\cite{BESIII:2019dme}. Even the interpretation of the $X(1750)$ remains uncertain, its role played in the $e^+ e^- \to \phi \pi^+ \pi^-$ reaction is also studied here.

This article is organized as follows. In Sec.~\ref{formalism}, we explain how to calculate the scattering amplitude and the total cross sections of $e^+ e^- \to \phi \pi^+\pi^-$ reaction. Numerical results and discussions are given in Sec.~\ref{results}, followed by a summary in the last section.

\section{Formalism} \label{formalism}

\begin{figure}[htbp]
\centering
\includegraphics[scale=0.52]{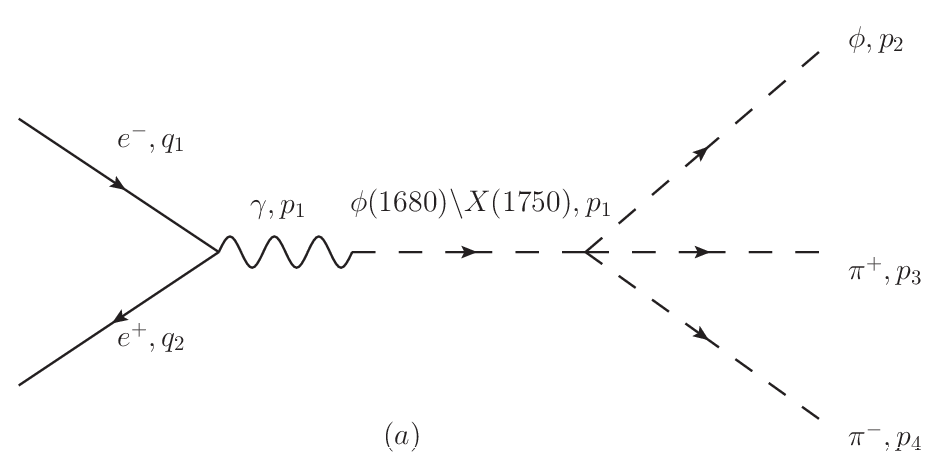}
\includegraphics[scale=0.52]{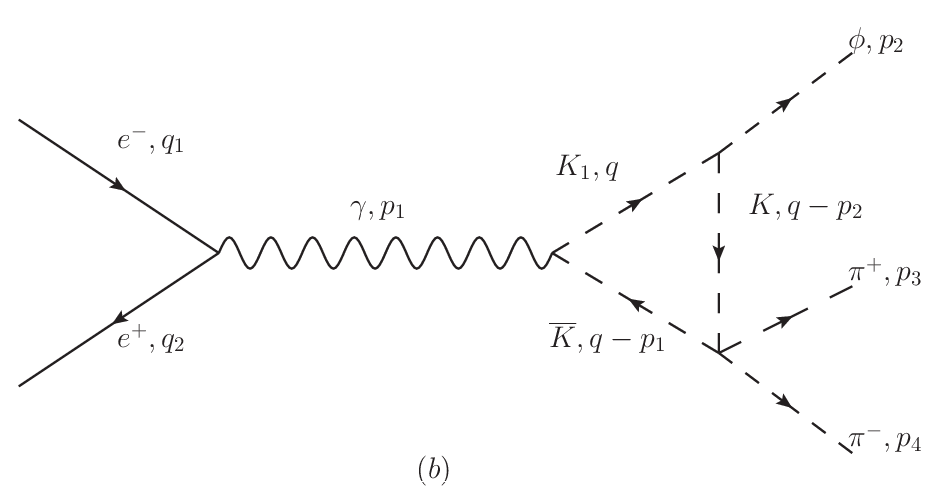}
\caption{Mechanisms for the process of $e^+e^-\rightarrow\phi\pi^+\pi^-$ involving the tree level diagrams [Fig.~\ref{fig:process} (a)] and the production of $K_1$ and $\bar{K}$ pair and then they transit to $\phi \pi^+\pi^-$ in the final state through the triangle diagram [Fig.~\ref{fig:process} (b)]. We also show the definition of the kinematical ($q_1, q_2, p_1, p_2, p_3, p_4$) that we use in the present calculations. }\label{fig:process}
\end{figure}

The reaction mechanism that we study for the $e^+ e^- \to \phi \pi^+ \pi^-$ reaction is shown in Fig.~\ref{fig:process}. First, there are two tree level diagrams [Fig.~\ref{fig:process} (a)], contributed by the $\phi(1680)$ meson and the $X(1750)$ meson which are created by the virtual photon via vector meson dominance model~\footnote{Note that there is no contribution from $\rho^0$ meson, because of the $C$-parity.}. Then, we considered also the triangle $K_1\mbox{-}K\mbox{-}\bar{K}$ intermediate state process [Fig.~\ref{fig:process} (b)]. That is a pair of $K_1\bar{K}$ mesons produced firstly through the one-photon exchange, and then the $K_1$ meson decays into $\phi$ and $K$ mesons in $S$-wave. Finally, the $K\bar{K}$ mesons produce the pair of $\pi^+$ and $\pi^-$ through the $S$-wave final state interactions, where the scalar meson $f_0(980)$ is dynamically generated~\cite{Oller:1997ti,Oller:1998zr,Gamermann:2006nm}. 

It is important to note that we will consider also the $\pi\pi \to \pi^+\pi^-$ final state interaction in Fig.~\ref{fig:process} (a). In addition, in the triangle loop depicted in Fig.~\ref{fig:process} (b), all three interaction vertices occur in the S-wave interaction, a feature that significantly enhances the contribution of the triangle singularity mechanism~\cite{Liu:2015taa,Bayar:2016ftu,Guo:2019twa}.

Then, the scattering amplitudes of $e^+ e^- \to \phi \pi^+ \pi^-$ reaction are given by:
\begin{widetext}
\begin{eqnarray}
\mathcal{M}_{1} &= & \frac{i e g_{R} M^2_{R}}{s [s - M_{R}^2+iM_{R}\Gamma_{R}]} \bar{v}_e(q_2,s_2) (\gamma^\alpha - \slash \!\!\! p_1 p^{\alpha}_1/s) u_e(q_1,s_1) \varepsilon_\alpha (p_2,s_\phi) (1+ G_{\pi\pi} \mathcal{M}_{\pi \pi \to \pi^+\pi^-}) , \label{Ma} 
\\
\mathcal{M}_{2} &= & \frac{i e \mathcal{F}(s) g_{\gamma K_1\bar{K}} g_{K_1\phi K}M_{K_1} }{s}  \int  d^4q   \frac{\bar{v}_e(q_2,s_2) (\gamma^\alpha - \slash \!\!\! q q^{\alpha}/q^2) u_e(q_1,s_1)  \varepsilon_\alpha (p_2,s_\phi) \mathcal{M}_{K\bar{K} \to \pi^+\pi^-} }{(q^2-M_{K_1}^2 + i M_{K_1}\Gamma_{K_1})[(q-p_2)^2 - m_K^2] [(q-p_1)^2 - m_K^2] }.
\label{Mb}
\end{eqnarray}
\end{widetext}
The subindices 1 and 2 stand for the diagrams of Figs.~\ref{fig:process} (a) and (b), respectively. In amplitude ${\cal M}_1$, we have considered the final state interactions of $\pi \pi \to \pi^+\pi^-$ in $S$-wave, where $G_{\pi\pi}$ is the loop function for $\pi^+\pi^-$ and $\pi^0 \pi^0$ channels~\cite{Oller:1997ti,Oller:1998zr}. In amplitude ${\cal M}_2$, the subscript $R$ represents the $\phi(1680)$ meson or the $X(1750)$ meson. The $\bar{v}_e(q_2,s_2)$ and $u_e(q_1,s_1)$ are Dirac spinors of $e^+$ and $e^-$, respectively, and $\varepsilon_\alpha (p_2,s_\phi)$ is the polarization vector of $\phi$ meson, with $s_1$, $s_2$, and $s_\phi$ their spin polarization variables. And $s = p^2_1$ is the invariant mass square of the initial $e^+ e^-$ system. The masses and widths for the $\phi(1680)$ and $X(1750)$ resonances are taken from the PDG~\cite{ParticleDataGroup:2024cfk}, with $M_{\phi(1680)} = 1680$ MeV and $\Gamma_{\phi(1680)} = 150$ MeV for the $\phi(1680)$, and $M_{X(1750)} = 1753.8$ MeV and $\Gamma_{X(1750)} = 120$ MeV for the $X(1750)$. Besides, $g_{\phi(1680)}$, $g_{X(1750)}$, $g_{\gamma K_1 \bar{K}}$, and $g_{K_1 \phi K}$ are effective coupling constants, which will be determined by fitting them to the experimental data. The $M_{K_1}$ and $\Gamma_{K_1}$ are the mass and width of the intermediate $K_1$ meson, and they are free parameters in this work.

In Eqs.~\eqref{Mb}, we include also a common used form factors $\mathcal{F}(s)$. Because there is no an unique theoretical way to introduce the form factor~\cite{Liu:1995st,Haberzettl:1998aqi,Davidson:2001rk,Janssen:2001wk}, we adopt here the scheme used in the previous works~\cite{Chen:2020xho,Wang:2021gle,Zhou:2022wwk,Chen:2025aes,Xie:2005sb,Xie:2007qt,Xie:2007vs,Cheng:2016hxi,Wang:2022vjm}. Here, we take the following parametrization for the form factor:
\begin{eqnarray}
\mathcal{F}(s) &=& e^{-b\sqrt{s}}, \label{ffs}
\end{eqnarray}
where $b$ is considered as a free model parameter, which can be determined by fitting experimental data.

\begin{figure}[htbp]
\centering
\includegraphics[scale=0.38]{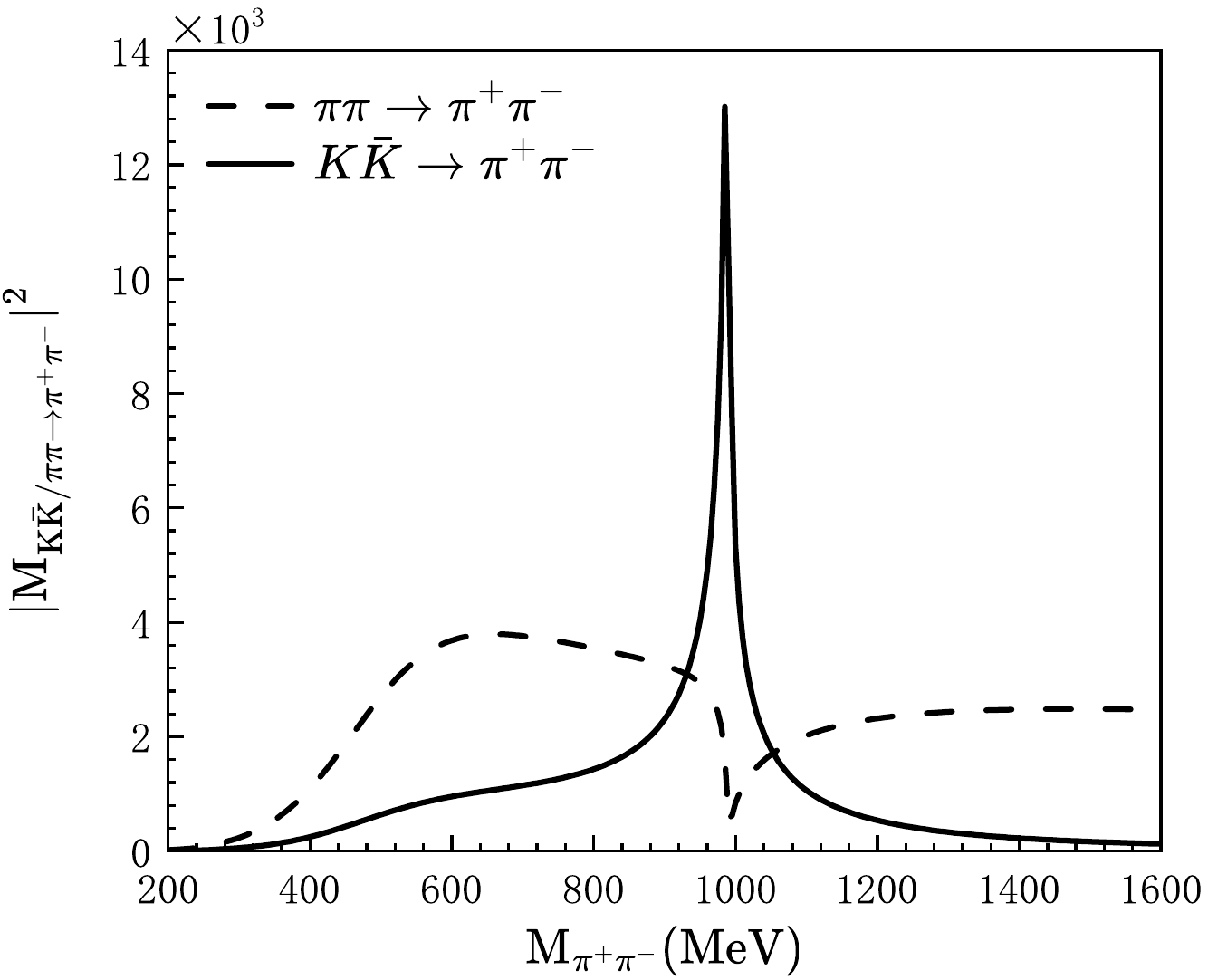}
\caption{The two-body scattering amplitudes of $K\bar{K} \to \pi^+ \pi^-$ and $\pi\pi \to \pi^+ \pi^-$ transition as a function of the invariant mass $M_{\pi^+ \pi^-}$.}\label{fig:mkk} 
\end{figure}

Then we turn to the two-body scattering amplitudes $\mathcal{M}_{ \pi \pi \to \pi^+\pi^-}$ and $\mathcal{M}_{ K\bar{K} \to \pi^+\pi^-}$, which stand for the final state interactions of $\pi \pi \to \pi^+ \pi^-$ and $K\bar{K} \to \pi^+ \pi^-$ transition in $S$-wave, respectively. These above two transition amplitudes depend on the invariant mass of $\pi^+\pi^-$ system, $M_{\pi^+\pi^-}$, and can be obtained by solving the Bethe-Salpeter equation as done in Refs.~\cite{Oller:1997ti,Oller:1998zr,Gamermann:2006nm}. In these final state interactions of $\pi\pi \to \pi^+\pi^-$ and $K\bar{K} \to \pi^+ \pi^-$, the scalar mesons $f_0(500)$ and $f_0(980)$ could be dynamically generated, as shown in Fig.~\ref{fig:mkk}. In the $\pi\pi \to \pi^+ \pi^-$ transition amplitude, there is also the signal for $f_0(980)$ resonance, but it is small compared with the strength of the signal for $f_0(500)$ resonance. Furthermore, it is a dip rather than a peak around the $K \bar{K}$ threshold, as discussed in Ref.~\cite{Dong:2020hxe}.

With the above scattering amplitudes, one can finally obtain the cross sections for the $e^+ e^- \to \phi \pi^+ \pi^-$ reaction, which reads,
\begin{eqnarray}
d\sigma (e^+ e^- \to \phi \pi^+ \pi^- ) &=&  \frac{|p_\phi| |p^*_\pi|}{32(\sqrt{s})^3(2\pi)^5} \times \nonumber \\
&\;& \hspace{-3.5cm} \sum_{s_1,s_2,s_\phi}|\mathcal{M}_{1}+\mathcal{M}_{2}|^2 dM_{\pi\pi} d\Omega_1  d\Omega^*_2,
\label{sigma}
\end{eqnarray}
where $M_{\pi \pi} = \sqrt{p^2_{34}}$ is the invariant mass of the $\pi^+ \pi^- $ system. $\Omega_1$ and $\Omega^*_2$ are the solid angles in the center-of-mass (CM) system of the $e^+ e^-$ collision and in the rest frame of the two-body $\pi^+ \pi^-$ final state, respectively. The $p_\phi$ and $p^*_\pi$ are the three-momenta of the $\phi$ meson and $\pi^+$ ($\pi^-$) meson in the CM frame and the $\pi^+\pi^-$ rest frame, respectively, which are given by
\begin{eqnarray}
|p_\phi|  &=&  \frac{\sqrt{(s-(m_\phi + M_{\pi \pi})^2)(s-(m_\phi - M_{\pi \pi})^2)}}{2\sqrt{s}}, \\
|p^*_\pi| &=& \frac{\sqrt{M^2_{\pi\pi} -4m^2_\pi}}{2}.
\end{eqnarray}

With the formalism and ingredients given above, the total cross section versus the center-of-mass energy $\sqrt{s}$ for the $e^+ e^- \to \phi \pi^+ \pi^-$ reaction is calculated by using a multi-particle phase space integration program, and then we compare the theoretical results to the experimental data. There are nine free parameters to be obtained by fitting to the experimental data: (1) two effective coupling constants $g_{R}$ for the production of $\phi(1680)$ or $X(1750)$ resonance related to the strength of the $\gamma$-$R$ and $R$-$\phi \pi\pi$ vertexes of Fig.~\ref{fig:process} (a); (2) two~\footnote{We consider two $K_1$ mesons in this work.} effective coupling constants $g_{K_1}$ for the triangle loop related to the strength of the $K_1 K \gamma$ and $ K_1 K \phi$ vertexes of Fig.~\ref{fig:process} (b); (3) two masses $M_{K_1}$ and two widths $\Gamma_{K_1}$ of the involved $K_1$ mesons; (4) parameter $b$ in the form factor ${\cal F}(s)$ for the direct coupling processes of Figs.~\ref{fig:process} (b).

\section{Numerical results and conclusions} \label{results}

We perform nine-parameter [$g_{\phi(1680)}$, $g_{X(1750)}$, $g^a_{K_1}$, $g^b_{K_1}$, $M^a_{K_1}$, $M^b_{K_1}$, $\Gamma^a_{K_1}$, $\Gamma^b_{K_1}$, and $b$] $\chi^2$ fits to the experimental data on the total cross sections of $e^+ e^- \to \phi \pi^+ \pi^-$ reaction from the BESIII~\cite{BESIII:2021aet} and BaBar~\cite{BaBar:2011btv} Collaborations. There are a total of 57 data points below energy $\sqrt{s} = 2600$ MeV. The obtained $\chi^2/{\rm dof}$ is 1.85, which is reasonably small. The fitted parameters are compiled in Table~\ref{tab:resultsoffit}.

\begin{table}[H]
\renewcommand\arraystretch{1.5}
\centering
\caption{The values of the model parameters determined by fitting them to the total cross sections of the $e^+ e^- \to \phi \pi^+ \pi^-$ reaction.}
\begin{tabular}{c|c|c|c}
\hline   \hline
Parameter & Fitted results & Parameter & Fitted results \\
\hline
$g_{\phi(1680)}(\times 10^{-2})$ & $9.3 \pm 1.7$  & $g^a_{ K_1}$ $(\times 10^{-2})$ & $9.8 \pm 1.1$\\
\hline
$g_{X(1750)}$ $(\times10^{-2})$ & $2.8 \pm 1.7$ & $g^b_{ K_1}$ $(\times 10^{-2})$ & $7.7 \pm 1.5$ \\
\hline
$M^a_{K_1}$ (MeV) & $1610 \pm 30$ & $\Gamma^a_{K_1}$ (MeV) & $41 \pm 11$ \\
\hline
$M^b_{K_1} $ (MeV)& $1895 \pm 50$ & $\Gamma^b_{K_1}$ (MeV)  & $32 \pm 15 $ \\
\hline
$b\ \mathrm{(GeV^{-1})}$  & $3.2\pm 0.1$ \\
\hline\hline
\end{tabular}
\label{tab:resultsoffit}
\end{table}

\begin{figure}[htbp]
\centering
\includegraphics[scale=0.38]{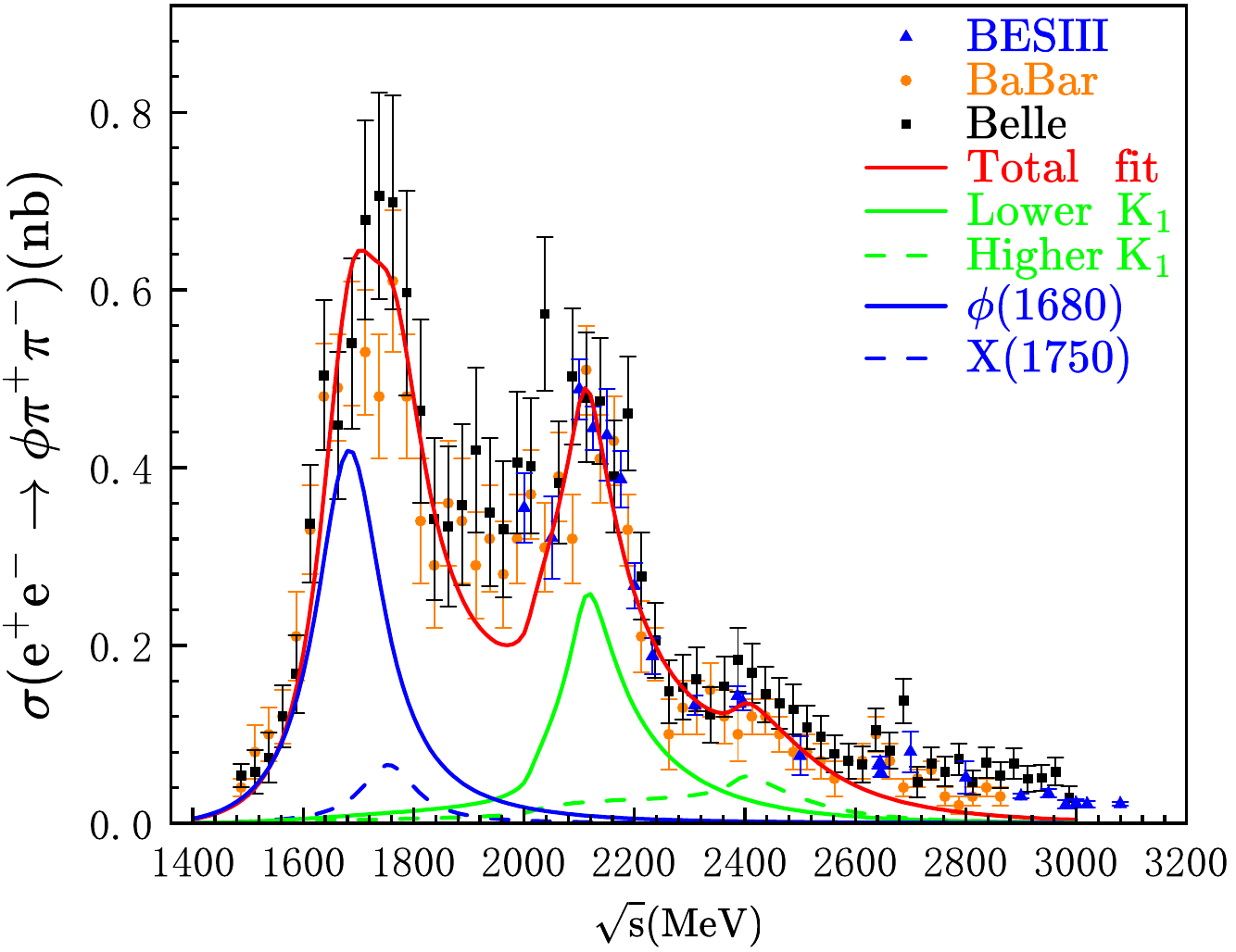}
\caption{Theoretical fitted results for the total cross sections of the $e^+e^- \to \phi \pi^+ \pi^-$ reaction. The experimental data are taken from Refs.~\cite{Belle:2008kuo,BaBar:2011btv,BESIII:2021aet}.}\label{fig:Total} 
\end{figure}

Using the fitted values of the model parameters, we calculate the total cross sections of the $e^+ e^- \to \phi \pi^+ \pi^-$ reaction, as shown in Fig.~\ref{fig:Total}. The theoretical results are compared with the experimental measurements from the Belle Collaboration~\cite{Belle:2008kuo}, BESIII Collaboration~\cite{BESIII:2021aet} and BaBar Collaboration~\cite{BaBar:2011btv}. Notably, we exclude the Belle Collaboration data~\cite{Belle:2008kuo} from our fitting due to their apparent inconsistency with the results from the other two experiments, particularly in the energy region between 2000 and 2100 MeV. The blue and green curves represent the contributions from tree diagrams [Fig.~\ref{fig:process} (a)], and triangle diagrams [Fig.~\ref{fig:process} (b)], respectively, while the red curve stands for the total contribution from all relevant diagrams. From Fig.~\ref{fig:Total} one can find that the peak of $\phi(2170)$ and the bump structure of $R(2400)$ can be well reproduced, thanks to contributions from the $K_1$-$K$-$\bar{K}$ triangle loops of Fig.~\ref{fig:process} (c). Within this framework, the $\phi(2170)$ [also the $R(2400)$] does not represent as a genuine state, but rather arises as an enhancement effect induced by the triangle singularity mechanism. This gives a new explanation for the nature of the $\phi(2170)$ state. 

The contribution of $\phi(1680)$ is predominant to the first peak structure, while the contribution of $X(1750)$ is small. However, its contribution is crucial to the total cross sections around the energy region from 1800 to 2000 MeV.

The fitted resonance parameters for the two $K_1$ mesons are: $M^a_{K_1} = 1610 \pm 30$ MeV, $\Gamma^a_{K_1} = 41 \pm 11$ MeV, $M^b_{K_1} = 1895 \pm 50$ MeV, and $\Gamma^b_{K_1} = 32 \pm 15$ MeV. According to the mass and width, the lower $K_1$ meson could be associated with the $K_1(1650)$ resonance as quoted in the PDG~\cite{ParticleDataGroup:2024cfk} within uncertainties. The only observed decay channels of $K_1(1650)$ resonance are $K \pi \pi$ and $K \phi$, which could be easily associated to the significant effective $K_1 K \phi$ coupling as obtained here. For the higher one, there is no $K_1$ meson around that energy region is reported in the PDG. However, a possible $K_1$ meson above $1.8$ GeV was predicted theoretically in Refs.~\cite{Garcia-Recio:2010enl,Garcia-Recio:2013uva}, which has the dynamically generated nature and with a large $K\phi$ decay mode.

On the other hand, the LHCb collaboration reported two excited $K_1$ states in the decay $B^+ \to J/\psi \phi K^+$~\cite{LHCb:2021uow}. The masses and widths of the two $K_1$ states are $M = 1861 \pm 10 ^{+16}_{-40} $ MeV, $\Gamma = 149 \pm 41 ^{+231}_{-23}$ MeV and $M = 1911 \pm 37 ^{+124}_{-48} $ MeV, $\Gamma = 276 \pm 50 ^{+319}_{-159}$ MeV, respectively. The parameters of the lower of these two states agree well with our heavier $K_1$ state.

\begin{figure}[htbp]
\centering
\includegraphics[scale=0.38]{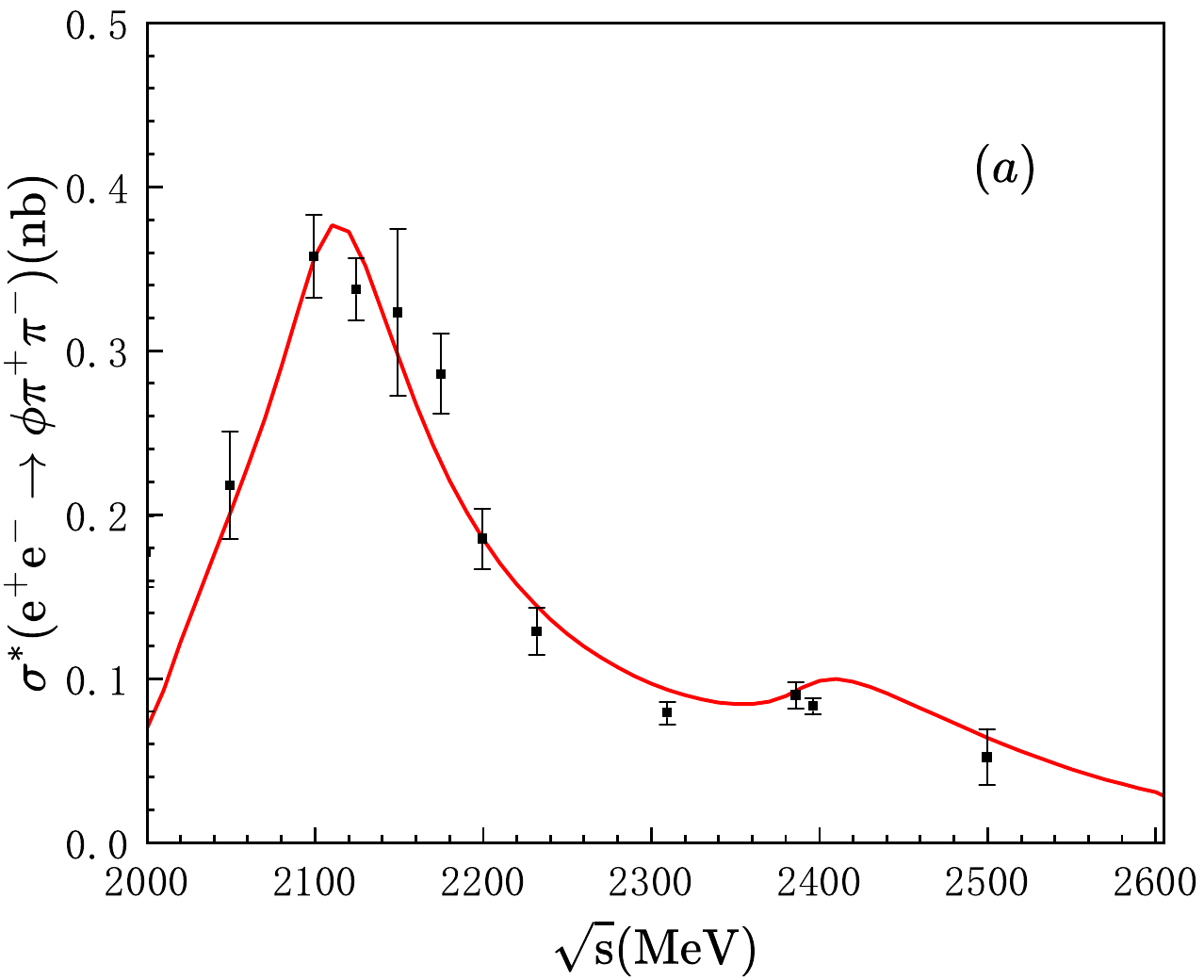}
\includegraphics[scale=0.36]{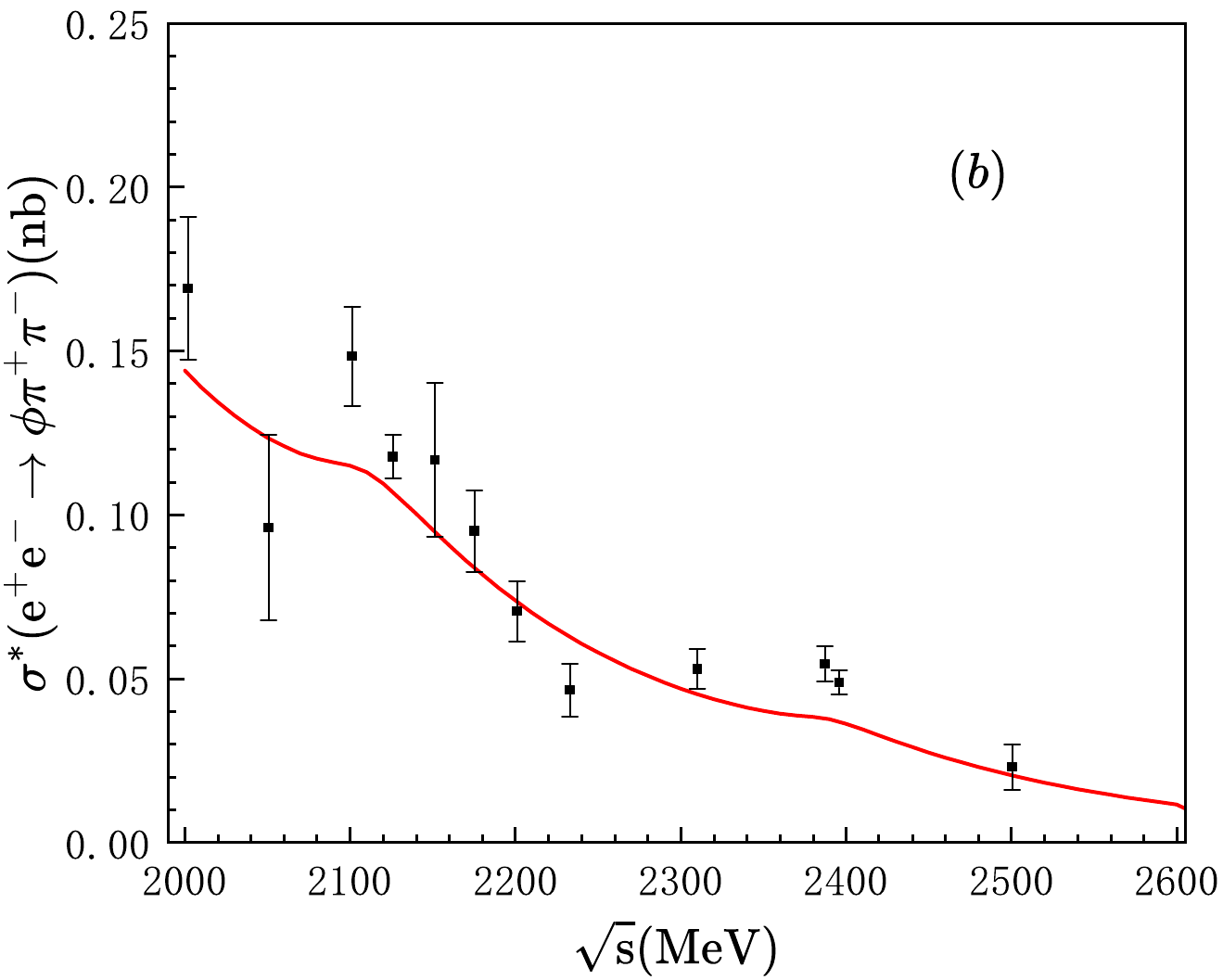}
\caption{The results for the cross sections of the process $e^+e^-\rightarrow\phi\pi^+\pi^-$ with the fitted parameters, in (a) $M_{\pi^+ \pi^-}\in [850,1100 ]$ MeV and (b) $M_{\pi^+\pi^-}\notin \left[850,1100\right]$ MeV. The experimental data are taken from Ref.~\cite{BESIII:2021aet}.}\label{fig:cut}
\end{figure}

Since the triangle singularity is much dependent on the kinematics and to further explore the contribution of the $e^+ e^- \to \phi f_0(980)$ for the production of $\phi(2170)$ state, we calculate the total cross sections of $e^+ e^- \to \phi \pi^+ \pi^-$ reaction in different energy region of $M_{\pi \pi }$, which was also done by the BESIII Collaboration in Ref.~\cite{BESIII:2021aet}. With fitted parameters shown in Table~\ref{tab:resultsoffit}, our theoretical results of the reduced total cross sections $\sigma^*(e^+ e^- \to \phi \pi^+ \pi^-)$ can be easily obtained and compared with the experimental data from Ref.~\cite{BESIII:2021aet} as shown in Fig.~\ref{fig:cut} by the red curves for $M_{\pi \pi} \in [850,1100]$ MeV and $M_{\pi \pi} \notin \left[ 850,1100 \right]$ MeV. It is found that we can describe well the experimental data. The $\phi(2170)$ is clearly seen in the energy region of $M_{\pi \pi} \in [850,1100]$ MeV. However, no significant $\phi(2170)$ signal is seen in the cross section line shape of $e^+ e^- \to \phi \pi^+ \pi^-$ reaction with $M_{\pi \pi} \notin \left[ 850,1100 \right]$ MeV. This is support the significant contribution of the subprocess $e^+ e^- \to \phi f_0(980)$ for the production of $\phi(2170)$ state in the $e^+ e^- \to \phi \pi^+ \pi^-$ reaction and also the production mechanism proposed here. Furthermore, the analysed total cross sections of $e^+ e^- \to \phi \pi^0 \pi^0$ reaction by the BESIII Collaboration~\cite{BESIII:2020vtu} can be also reproduced by replacing $\pi^+ \pi^-$ with $\pi^0 \pi^0$.

On the other hand, for the elastic rescattering process, $e^+ e^- \to \phi K^+ K^-$ reaction, in addition to the triangle diagram, there also be corresponding intermediate $K_1$ mesons production in the tree diagram.~\footnote{Until now, there is no $Z_s$ state which was observed in the $\phi \pi$ channel.} When it is added coherently to the triangle rescattering diagram, the effect of the triangle diagram is nothing more than multiplying a partial wave amplitude of the tree diagram by a phase factor, and therefore the singularities of the triangle diagram cannot produce obvious peaks in the total cross sections of the $e^+ e^- \to \phi K^+ K^-$ reaction~\cite{Guo:2019twa,Anisovich:1995ab,Debastiani:2018xoi}. However, the $e^+ e^- \to \phi K^+K^-$ reaction with the $K^+K^-$ rescattering to a$\pi^+\pi^-$ pair plays an important role in the $e^+ e^- \to \pi^+\pi^-$ reaction.

However, further experimental investigations about the $\phi K$ invariant mass distributions in the $e^+ e^- \to \phi K^+ K^-$ reaction could provide valuable insights into the excited $K_1$ states, which currently remain poorly understood due to limited experimental data and theoretical studies~\cite{ParticleDataGroup:2024cfk,Garcia-Recio:2010enl,Garcia-Recio:2013uva,Roca:2005nm}. A comprehensive analysis of the $\phi K$ distributions would significantly enhance our understanding of the properties and decay mechanisms of these exotic meson states when higher statistics data are available.

\section{Summary}

We have investigated the production mechanism of $\phi(2170)$ in the $e^+e^- \to \phi \pi^+ \pi^-$ reaction through two distinct triangle loop diagrams involving $K_1$, $\bar{K}$ and $K$ mesons. It is found that when the masses of $K_1$ are taken as $1610$ MeV and $1895$ MeV respectively, one can get a good fitting and provide agreement with the recent BESIII measurements for the production of both $\phi(2170)$ and $R(2400)$.

The study here demonstrates that the $\phi(2170)$ signal can be fully accounted for by the triangular singularity mechanism mediated by a $K_1$ resonace with mass about 1610 MeV. More intriguingly, the fitting results for $R(2400)$ suggest the existence of a previously unobserved $K_1$ meson with quantum number $I(J^P)=\frac{1}{2}(1^+)$ and a mass around 1895 MeV. We strongly encourage future experimental efforts to search for this predicted $K_1$ state, which would provide crucial validation of our theoretical framework.

\begin{acknowledgments}

We would like to thank Profs. Eulogio Oset and Alberto Martinez Torres for careful reading the article and for constructive comments. Useful discussions with Prof. Wen-Biao Yan is also appreciated. This work is partly supported by the National Key R\&D Program of China under Grant No. 2023YFA1606703, and by the National Natural Science Foundation of China under Grant Nos. 12435007, 12361141819, 11975165 and 12235018.

\end{acknowledgments}

\normalem
\bibliographystyle{apsrev4-1.bst}
\bibliography{reference.bib}

\end{document}